\documentclass[aps,prd,preprint,11pt,superscriptaddress,nofootinbib,floatfix]{revtex4-1}
\usepackage{fancyhdr}
\usepackage{amsfonts}
\usepackage[latin1]{inputenc}
\usepackage{graphicx}
\usepackage{mathrsfs}
\usepackage{amssymb}
\usepackage{amsmath}
\usepackage{verbatim}
\usepackage{multirow}
\usepackage{soul}
\usepackage{mciteplus}
\usepackage{subfigure}
\usepackage[usenames,dvipsnames]{color}
\usepackage{tabularx}
\usepackage[framemethod=TikZ]{mdframed}
\mdfdefinestyle{MyFrame}{%
    linecolor=gray,
    outerlinewidth=0.4pt,
    roundcorner=4pt,
    innertopmargin=0.5\baselineskip,
    innerbottommargin=0.5\baselineskip,
    innerrightmargin=\parindent,
    innerleftmargin=\parindent,
    backgroundcolor=gray!05!white}
\newcommand{\subroutine}[2]{\\ \begin{mdframed}[style=MyFrame]\small\begin{tabularx}{\linewidth}{r@{}X} \texttt{#1} & \texttt{#2} \end{tabularx}\end{mdframed}}

\graphicspath{{figs/}}

\newcommand{\HB}{\texttt{Higgs\-Bounds}}
\newcommand{\HBv}[1]{\texttt{HiggsBounds-#1}}
\newcommand{\HS}{\texttt{HiggsSignals}}
\newcommand{\CP}{\mathcal{CP}}
\newcommand{\CL}{~\mathrm{C.L.}}

\include{paperdef}

\begin{document}

%\date{\today}
\title{Applying Exclusion Likelihoods from LHC Searches to Extended Higgs Sectors}
\vspace*{1.0truecm}
\author{Philip Bechtle}
\email{bechtle@physik.uni-bonn.de}
\affiliation{Physikalisches Institut der Universit\"at Bonn, Nu{\ss}allee 12, 53115 Bonn, Germany}
\author{Sven Heinemeyer}
\email{Sven.Heinemeyer@cern.ch}
\affiliation{Instituto de F\'isica de Cantabria (CSIC-UC), Santander, Spain}
\author{Oscar St{\aa}l}
\email{oscar.stal@fysik.su.se}\
\affiliation{The Oskar Klein Centre, Department of Physics, Stockholm University, SE-106 91 Stockholm, Sweden}
\author{Tim Stefaniak}
\email{tistefan@ucsc.edu}
\affiliation{Santa Cruz Institute for Particle Physics (SCIPP), University of California, Santa Cruz, CA 95064, USA}
\author{Georg Weiglein\vspace{0.3cm}}
\email{Georg.Weiglein@desy.de}
\affiliation{Deutsches Elektronen-Synchrotron DESY, Notkestra{\ss}e 85, 22607 Hamburg, Germany\vspace{0.5cm}}

\renewcommand{\abstractname}{\vspace{0.2cm} Abstract}

\begin{abstract}
\vspace{0.5cm}
LHC searches for non-standard Higgs bosons decaying into tau lepton
pairs constitute a sensitive experimental probe for physics beyond the
Standard Model (BSM), such as Supersymmetry (SUSY). Recently, the
limits obtained from these searches have been presented by the CMS
collaboration in a nearly model-independent fashion --- as a narrow
resonance model --- based on the full $8\tev$ dataset. In addition to
publishing a $95\%~\mathrm{C.L.}$ exclusion limit, the full likelihood
information for the narrow resonance model has been released. This provides valuable information that
can be incorporated into global BSM fits. We present a simple algorithm
that maps an arbitrary model with multiple neutral Higgs bosons onto the
narrow resonance model and derives the corresponding value for the
exclusion likelihood from the CMS search. This procedure has been
implemented into the public computer code \HB\ (version \texttt{4.2.0}
and higher). We validate our implementation by cross-checking
against the official CMS exclusion contours in three Higgs benchmark scenarios in the Minimal Supersymmetric Standard Model (MSSM), and find very good agreement. Going beyond validation, we discuss the combined constraints of the $\tau\tau$ search and the rate measurements of the SM-like Higgs at $125\gev$ in a recently proposed MSSM benchmark scenario, where the lightest Higgs boson obtains SM-like couplings independently of the decoupling of the heavier Higgs states. Technical details for how to access the likelihood information within \HB\ are given in the appendix. The program is available at \url{http://higgsbounds.hepforge.org}.

\end{abstract}

\preprint{BONN-TH-2015-08, DESY 15-093, SCIPP 15/05}

\maketitle

%\clearpage

\newpage

%%%%%%%%%%%%%%%%%%%%%%%%%%%%%%%%%%%%%%%%%%%%%%%%%%%%%%%%%%%%%%%%%%%%%%%%%%%%%%%
%%%%%%%%%%%%%%%%%%%%%%%%%%%%%%%%%%%%%%%%%%%%%%%%%%%%%%%%%%%%%%%%%%%%%%%%%%%%%%%

\section{Introduction}

The search for Higgs bosons~\cite{Englert:1964et,*Higgs:1964ia,*Higgs:1964pj,*Guralnik:1964eu,*Higgs:1966ev,*Kibble:1967sv}
continues to be a cornerstone of the physics program at the Large Hadron Collider (LHC). After the discovery of a Higgs
boson by ATLAS~\cite{Aad:2012tfa} and CMS~\cite{Chatrchyan:2012ufa} it is crucial to find out whether the detected particle is part of a Higgs sector that contains several physical states. Higgs sectors of this kind are predicted in many theories of physics beyond the Standard Model (SM). For the understanding of the mechanism of electroweak symmetry breaking 
two complementary experimental endeavors are important: On the one hand the precise determination of the properties of the Higgs signal detected at around $125 \gev$, and on the other hand the search for additional Higgs bosons. Both are crucial
 in the quest to identify the underlying physics. The existing limits 
from the Higgs searches at LEP, the Tevatron and the LHC 
already put very important constraints on the parameter spaces of
different models that provide a Higgs-like state compatible with the detected signal.
More data on both the detected signal and on searches for additional Higgs bosons will further enhance the sensitivity for
discriminating possible scenarios of new physics from the SM and from each other. 

In order to facilitate the available experimental information from the
Higgs searches at LEP, the Tevatron and the LHC, expressed in terms of
relatively model-independent cross-section limits for testing a wide
variety of theoretical models, the program 
\HB~\cite{Bechtle:2008jh,Bechtle:2011sb,Bechtle:2013gu,Bechtle:2013wla} 
has been developed. Experimental information on the Higgs signal detected at
a mass value of around $125\gev$ is utilized in the sister program 
\HS~\cite{Bechtle:2013xfa} for testing the theoretical predictions from any kind of Higgs sector. 
The experimental information on the detected signal
incorporated in \HS\ is turned into a $\chi^2$ likelihood, which is suitable for
the inclusion into global fits (see, e.g., \citeres{deVries:2015hva,Bechtle:2012jw,Bechtle:2012zk,
Bechtle:2013mda,Bechtle:2014yna, Bechtle:2014ewa}), where in addition many other observables are taken into account.
In contrast, exclusion limits have traditionally been presented 
in terms of $95\%\CL$ limits, which a priori only provide the
information whether a particular parameter point is excluded or not at
the $95\%\CL$ by the considered search channel. In a global fit, where
the predictions of a model are confronted with a large number of
observables, it would usually be too restrictive to disregard a certain
parameter point just because it falls outside of the $95\%\CL$ region of a
single search channel. In fact, testing a large variety of observables
one would expect that the measured values of some observables lie
outside of the respective $95\%\CL$ regions for purely statistical
reasons. It would therefore be very desirable if also negative experimental outcomes from Higgs searches were 
provided in terms of the likelihood information in the relevant
parameters, instead of a simple binary rejection or
acceptance at a certain confidence level (C.L.). Up to now, likelihood information was available in \HB\
only for the results from the LEP Higgs searches~\cite{Bechtle:2013wla}, 
while for all search channels at the Tevatron and LHC only $95\%\CL$ limits have been accessible.
We report here on significant progress in this direction for the LHC Higgs
boson search in the $\tau^+\tau^-$ final state, which plays a central
role in the search for additional Higgs bosons.

Many models that can accommodate a SM-like Higgs boson at $125\gev$, such as the Minimal
Supersymmetric Standard Model (MSSM) or the various types of Two-Higgs-Doublet Models (2HDM), predict additional Higgs bosons that decay predominantly into SM fermions. Therefore, LHC searches for new neutral Higgs
bosons decaying to  $\tau^+\tau^-$ play a crucial role. In particular within the MSSM, these searches lead to large excluded regions in the parameter space. The highest experimental sensitivity occurs for smaller values of the $\cp$-odd Higgs boson mass,
$\MA$, and larger values of $\tb$, the ratio of the two vacuum expectation values~\cite{Khachatryan:2014wca,Aad:2014vgg}.

One complication that arises for this search channel is the fact that two different production modes, gluon fusion and $b$~quark associated production, can both be important. Their individual contributions to the signal rate can vary strongly over the parameter
space. Since the acceptances of these two channels can also be very different, a two-dimensional cross section interpretation 
for the $\tau^+\tau^-$ final state is desirable as a basis for (close to) model-independent exclusion limits.
Recently, the CMS collaboration has published the likelihood information for their Higgs boson search in the $\tau^+\tau^-$ final
state~\cite{Khachatryan:2014wca}. The likelihood is given as a function of the two relevant Higgs production channels, gluon fusion and $b$~quark associated production, for various mass values of the narrow resonance assumed for the signal model.

In this paper we investigate the application of this new experimental information for testing the theoretical predictions of extended Higgs sectors and its incorporation in global fits. We develop a simple algorithm that maps an arbitrary model with in general several neutral Higgs bosons onto the narrow resonance model. In this way the corresponding value of the exclusion likelihood from the CMS search for the tested model can be determined. We furthermore describe the inclusion of this likelihood information
into the publicly available Fortran code \HB~\cite{Bechtle:2008jh,Bechtle:2011sb,Bechtle:2013gu,Bechtle:2013wla}.
For nearly any model under consideration, \HB\ provides an evaluation of the exclusion likelihood for a model parameter point
based on the information from \citere{Khachatryan:2014wca}. While the new likelihood information goes well beyond the standard test whether a particular parameter point is excluded at the $95\%\CL$, the likelihood information can also be employed to run \HB\ in this ``standard'' mode. In this case, \HB\ determines the parameter region that is excluded at the $95\%\CL$ based on {\em all} available searches, including the new $\tau^+\tau^-$ result from CMS.
The new version of \HB\ can be used together with its sister program \HS~\cite{Bechtle:2013xfa} in order to take into account both the
information from search limits and from the detected signal for a comprehensive test of Higgs phenomenology.
Both codes are available at: 
\begin{center}\url{http://higgsbounds.hepforge.org}\end{center}

The paper is organized as follows. In \refse{sec:expres} we briefly summarize the experimental results that are used as input for our investigations. Details of the employed algorithm and the implementation of the exclusion likelihood of \citere{Khachatryan:2014wca} into \HB\ are given in \refse{Sec:algorithm}. The validation in various MSSM Higgs benchmark scenarios is discussed in \refse{sec:validation}. As an application, in \refse{sec:application} we investigate the constraints on a certain benchmark scenario in the MSSM that are obtained from using the new exclusion likelihood in combination with the information on the detected signal incorporated in \HS. We conclude in \refse{sec:conclusions}. Finally, all relevant information needed to run \HB\ to obtain the likelihood information for the $\tau^+\tau^-$ Higgs search channel for any parameter point under investigation are contained in an Appendix.

%%%%%%%%%%%%%%%%%%%%%%%%%%%%%%%%%%%%%%%%%%%%%%%%%%%%%%%%%%%%%%%%%%%%%%%%%%%%%%%
%%%%%%%%%%%%%%%%%%%%%%%%%%%%%%%%%%%%%%%%%%%%%%%%%%%%%%%%%%%%%%%%%%%%%%%%%%%%%%%

\newpage
\section{Experimental results}
\label{sec:expres}

This section briefly summarizes the experimental results from the CMS non-standard Higgs search in the $\tau\tau$ final state~\cite{Khachatryan:2014wca}, that we have used as starting
point for our investigation and that we have implemented in \HB. The
search analysis of CMS is carried out in two separate selection categories: One
requiring the presence of at least one $b$-tagged jet, and one without the presence of a
$b$-tag. The former category is enriched by the production of a Higgs
boson, denoted generically by $\phi$, in association with two $b$
quarks, $gg\to b\bar{b}\phi$, while the latter is dominated by the gluon fusion process,
$gg\to\phi$. Hence, the search features sensitivity to the two different
production modes separately, which enables the presentation of the
search results in terms of individual signal strengths in both
production modes for all tested Higgs boson masses. 
Separate information on the two production modes is 
an indispensable ingredient for enabling the presentation of (close to) model-independent exclusion limits or measurements, in case of a
discovery.

The data is further classified in categories defined by the two $\tau$ lepton decay modes: $e\tau_h$, $\mu\tau_h$, $e\mu$, $\mu\mu$ and $\tau_h\tau_h$, where $\tau_h$ denotes a hadronically decaying $\tau$ lepton. Using a maximum likelihood technique, an estimator for the true $\tau\tau$ invariant mass, $m_{\tau\tau}$,
  is reconstructed from the momenta of the visible $\tau$ decay products and
  the missing transverse energy in the event. The uncertainty of the $m_{\tau\tau}$ reconstruction is estimated to be around $20\%$ when averaged over all
  decay modes~\cite{Khachatryan:2014wca}.

  The resulting $m_{\tau\tau}$ spectrum in all categories
  ($b$-tag and $\tau$ decay) separately is then subject to a profile
  likelihood analysis~\cite{CMS-NOTE-2011-005}, where the background
  parametrization, obtained from control region data and Monte Carlo
  simulation, and the signal shape parametrization are fitted simultaneously to the
  reconstructed mass spectrum. The fit is performed individually for
  test masses $m_{\phi}$ between $90\gev$ and $1\tev$, and the results are
  interpolated between the test masses.

CMS interprets the results both in a nearly model-independent
    way\footnote{The presentation of the search results in terms of a limit on
the inclusive total cross section times branching ratio inevitably involves
a slight model dependence from the extrapolation to the inclusive
quantity. In other words, the expectation of the kinematic distributions of the signal and/or background is model dependent.}
 for a single narrow resonance $\phi$, and, in a
    model-specific context, in the MSSM, where three neutral Higgs
  bosons $h$, $H$ and $A$ potentially comprise the signal.  The latter
  interpretation performs a likelihood ratio hypothesis test
for the two hypotheses of a single SM-like Higgs boson at $m_h=125\gev$ 
with exact SM properties and, alternatively, for the signal consisting of all
three neutral Higgs bosons of the MSSM. %This offers the advantage to combine 
In the latter case, the $m_{\tau\tau}$ distributions of the $h/H/A\to\tau\tau$ decays are combined \emph{before} the calculation of the likelihood. Note that, whereas the model-specific limits for the MSSM are based on the full integrated luminosity of the combined $7+8\tev$ dataset, the results for the single narrow resonance model are obtained from only the $8\tev$ dataset.
  
  Since \HB\ is designed to test \emph{any} extended Higgs sector, with
\emph{any} coupling properties of the $125\gev$ Higgs candidate (if the
model under consideration provides such a candidate; obviously the
phenomenological interest in other models is rather limited) and \emph{any}
masses and properties of the remaining Higgs spectrum, the
nearly model-independent single resonance results are chosen for the implementation
in \HB. On the one hand, this is the only possibility unless one is
willing to adopt further model-dependent assumptions. On the other
hand, it will in general yield weaker, i.e.~more conservative, limits
than a dedicated analysis taking into account the full structure of the
considered model. For example, in the MSSM this will yield a conservative limit whenever either
$m_{H/A}\approx m_h$, or, more generally, whenever the model predicts that
more than one Higgs boson have a non-negligible signal yield and contribute
in \emph{different} regions in $m_{\tau\tau}$. Since the likelihood is constructed for
  single resonances, such a case cannot be properly reconstructed from the
  likelihood. However, if e.g.~the heavy MSSM Higgs bosons $H$ and $A$ contribute at the
  same point in $m_{\tau\tau}$, their signal rates can be added before
  interpreting the likelihood. In this case the implementation is not necessarily
  conservative. A detailed study on the applicability of these limits to the MSSM benchmarks is presented in Section~\ref{sec:validation}.
  
  The profile likelihood analysis follows the standard implementation:
  The test statistics is given by 
  \begin{align}
    q_\mu = -2 \ln \frac{\mathcal{L}(N | \mu \cdot s(m) + b, \hat{\theta}_\mu)}{\mathcal{L}(N | \hat{\mu} \cdot s(m) + b, \hat{\theta} )}.
   \label{Eq:teststat}
  \end{align}
Here $N$ is the number of measured events, $b$ and $s(m)$ the number of expected background
  and signal events for a given resonance mass hypothesis $m$, $\mu$
  the signal strength modifier, and $\theta$ are the nuisance
  parameters decribing the systematic uncertainties. $\hat\theta_\mu$
  maximizes the likelihood in the numerator given a certain value of
  $\mu$, whereas the likelihood reaches its global maximum at
  $\hat\mu$ and $\hat\theta$, which is given in the denominator. 
  The constraint $0 \le \hat{\mu} \le \mu$ is employed to not
  penalize the model for a possible excess of the data over the signal
  plus background prediction. It should be noted
  that the signal yield contains two independent components, corresponding to the two production modes
    $gg\to \phi$ and $gg\to b\bar{b}\phi$. Thus, $\mu$ and $s(m)$ are two-component vectors.

  Using toy Monte Carlo techniques or asymptotic expressions for large
  statistics, the expected probability distributions
  $P(q_{\mu}|\text{hypothesis})$ can be constructed for the test
  statistics given above. In the model-independent analysis used
  here, the hypothesis either consists of $H_1=\mu \cdot s(m)+b$ for
  the case of the presence of a single narrow resonance with a given
  mass $m$ and signal yield $\mu \cdot s(m)$, and of $H_2=b$ for
  the background. Using these hypothesis definitions and the
  observed value of the likelihood ratio, $q_{\mu}^\text{obs}$, given by Eq.~\eqref{Eq:teststat} with $N$ given by the actual observed number of events, $N=N_\text{obs}$, the likelihood
  ratio technique can be used to define the $\mathrm{CL}_s$ as
  \begin{align}
    \mathrm{CL}_s(\mu) = \frac{P(q_\mu \ge q_\mu^\mathrm{obs} | \mu \cdot s(m) + b)}{P(q_\mu \ge q_\mu^\mathrm{obs} | b)}.
  \end{align}
  In a stand-alone search, the criterion $\mathrm{CL}_s \le \alpha$ is
  then used to exclude the presence of a signal at $1-\alpha$
  confidence level (C.L.). For the expected limit, the observed data
  in the calculation of $q_{\mu}^\text{obs}$ is replaced by the median of
  the background-only expectation for $q_{\mu}$. A model-independent
  limit, e.g. at the $95\%\CL$, can then be derived by varying $\mu$
  until $\mathrm{CL}_s=0.05$. The value $\mu$ at which this happens then
  represents the signal strength modifier which is just allowed at the
  $95\%\CL$.

For any given model, \HB\ reconstructs the predicted signal yield $s(m)$
from the theoretical input provided by the user, and obtains the
corresponding value of the test statistics $q_{\mu=1}$ (or simply denoted
$q_{\rm model}$) from the CMS likelihood data. The details of this
procedure will be described in the following section. Note that \HB\
directly employs the expectation and observation of the test statistics,
$q_\mu^\text{exp}$ and $q_\mu^\text{obs}$, respectively, as the provided CMS
data does not allow for a full reconstruction of the $\mathrm{CL}_s$ value.
Nevertheless, in the limit of large numbers, the test statistics can be
approximated by a chi-squared differences function above minimum, $q_\mu
\approx \Delta\chi^2 = \chi^2 - \chi^2_\text{min}$, as
\begin{align}
\chi^2\, &{\approx} - 2\ln \mathcal{L}(N | \mu \cdot s(m) + b, \hat{\theta}_\mu),\\
\chi^2_\text{min}\, &{\approx} - 2\ln \mathcal{L}(N | \hat{\mu} \cdot s(m) + b, \hat{\theta}).
\end{align}
Thus, for example, we approximately obtain the two-dimensional limit at the $68\%$ and $95\%\CL$ when the test statistics $q_\mu$ takes the values $2.28$ and $5.99$, respectively.
  
\begin{figure}[h]
\centering
\subfigure[\label{fig:CMS_BG_m125}~Observed exclusion likelihood, $q_\mu^\text{obs}$, and approximated exclusion contours for a resonance mass of $m_\phi = 125\gev$.]{\includegraphics[width=0.48\textwidth]{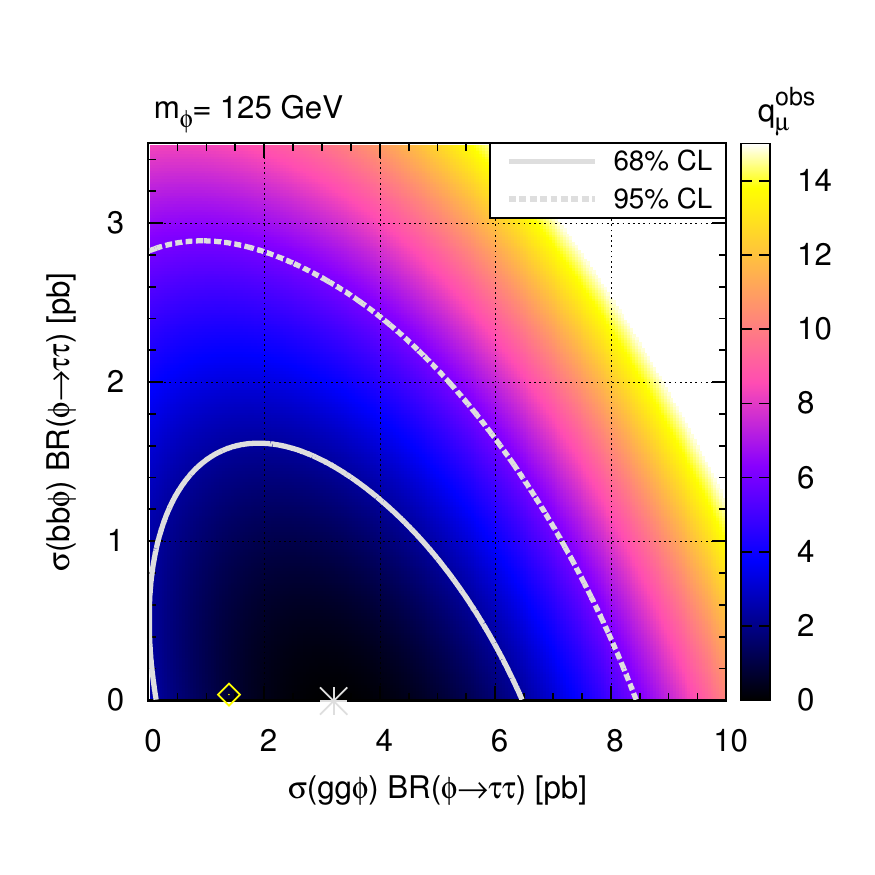}}
\hfill
\subfigure[\label{fig:CMS_BG_m300}~Observed exclusion likelihood, $q_\mu^\text{obs}$, and approximated exclusion contours  for a resonance mass of $m_\phi = 300\gev$.]{\includegraphics[width=0.48\textwidth]{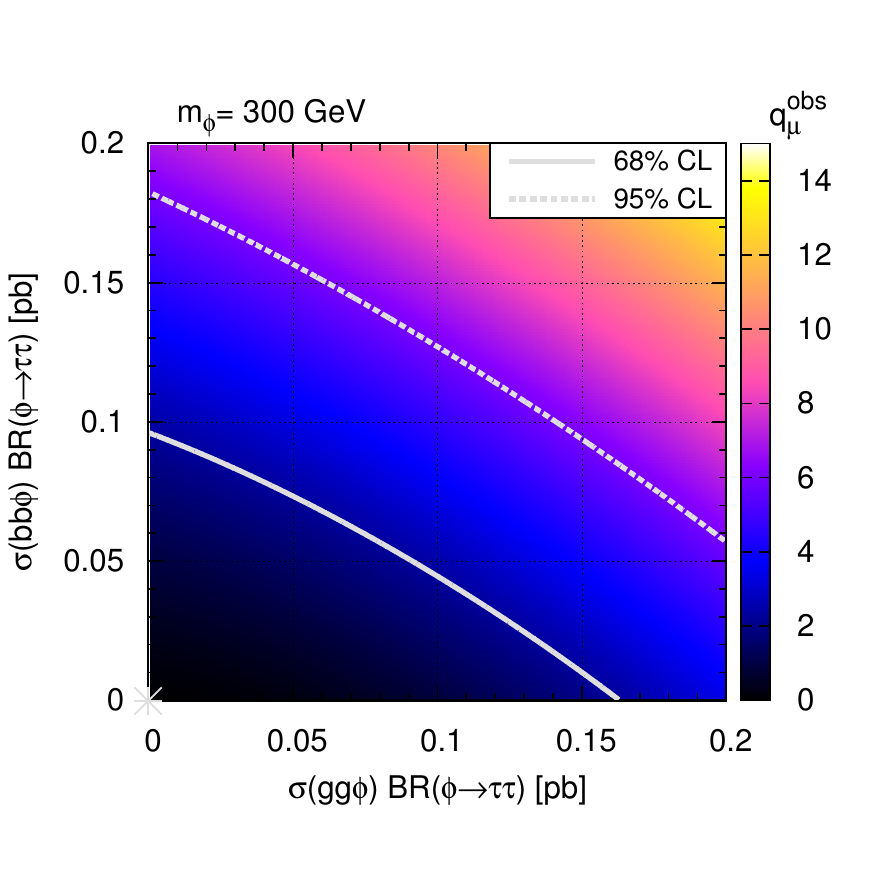}}
\caption{Results for the observed exclusion likelihood, $q_\mu^\text{obs}$, from the CMS
  $\phi\to\tau\tau$ analysis~\cite{Khachatryan:2014wca}, assuming a narrow
  resonance mass, $m_\phi$, of $125\gev$ (\emph{a}) and $300\gev$
  (\emph{b}). The solid (dashed) lines are obtained at $q_{\mu}^\text{obs}~=~2.28~(5.99)$ and indicate the approximate $68\%~(95\%)\CL$ allowed regions of a Higgs boson signal. The gray asterisk indicates the location of the global maximum of the likelihood. In (\emph{a}) the yellow hollow diamond indicates the prediction of a Higgs boson at $125\gev$ with SM signal strength.
}
\label{fig:CMSresults} 
\end{figure}

As an example, we show the observed likelihood distribution, $q_{\mu}^\text{obs}$, for test
masses of $125\gev$ and $300\gev$ in Fig.~\ref{fig:CMSresults}. We also
indicate the approximated $68\%\CL$ and $95\%\CL$ limits by contour lines. It
should be kept in mind that these contours are only for illustrational purposes. In \HB\ the full likelihood information $q_\mu$ is used, and the specific limit at a certain C.L.~can be easily obtained from this information.

The implementation of the likelihood for the CMS $\phi\to\tau\tau$ search differs in two significant ways from the implementation of
the LEP Higgs search $\chi^2$ in \HB, which is already available since version 2.0.0~\cite{Bechtle:2011sb}. In the LEP   implementation, each Higgs search channel, comprised of \emph{one} production mode and \emph{one} Higgs boson decay mode, is treated separately, thus no combination of production modes is applied or possible for the user. In addition, in the LEP implementation the $\chi^2$ is estimated from the $\text{CL}_{s+b}$ value in each channel at the given signal strength prediction using Gaussian approximations. In contrast, for the CMS $\phi\to\tau\tau$ search the exact values of the test statistics $q_{\mu}$ as presented by CMS are used and properly combined for both production modes.

%%%%%%%%%%%%%%%%%% F I G U R E %%%%%%%%%%%%%%%%%%%%%%%%%%%%%%%%%%%%%%%%%%%%%%%%
%%%%%%%%%%%%%%%%%% F I G U R E %%%%%%%%%%%%%%%%%%%%%%%%%%%%%%%%%%%%%%%%%%%%%%%%

%%%%%%%%%%%%%%%%%%%%%%%%%%%%%%%%%%%%%%%%%%%%%%%%%%%%%%%%%%%%%%%%%%%%%%%%%%%%%%%
%%%%%%%%%%%%%%%%%%%%%%%%%%%%%%%%%%%%%%%%%%%%%%%%%%%%%%%%%%%%%%%%%%%%%%%%%%%%%%%

%\newpage
\section{Likelihood reconstruction for extended Higgs sectors}
\label{Sec:algorithm}
For the construction of the exclusion likelihood from the $H\to\tau\tau$ search, we make use of
the following quantities: For each neutral Higgs boson,
$h_i$~($i=1,\dots,N$), in a model with $N$ neutral Higgs bosons, we have a
prediction of the mass, $m_i$ (where the relevant range is  currently $m_i \in [90,~1000]\gev$), the gluon fusion production cross section, $\sigma(gg\to h_i)$, the cross section for production in association with $b$ quarks, $\sigma(gg\to b\bar{b}h_i)$, and the branching fraction $\mathrm{BR}(h_i\to \tau\tau)$.

The main algorithm for the likelihood reconstruction proceeds as follows:
\begin{enumerate}
\item Signal rates of multiple Higgs bosons that cannot be resolved by the experimental analysis are added. We thus combine the signal predictions for two Higgs boson $h_i$ and $h_j$~$(j \ne i)$, if
\begin{align}
|m_i - m_j| \le 20\% \cdot \max(m_i, m_j).
\label{Eq:combination} 
\end{align}
Each Higgs boson can appear in different such combinations. For each combination, also called \emph{Higgs cluster} and labeled with capital characters in the following, we evaluate the physical quantities as follows: We assume that the total rates are given by the incoherent sum of the signal rates of the individual Higgs bosons in the cluster,
\begin{align}
\sigma(gg\to h_I \to \tau\tau) &= \sum_k \sigma(gg\to h_k) \cdot \mathrm{BR}(h_k\to \tau\tau), \label{Eq:cluster_ggH}\\
\sigma(gg\to b\bar{b} h_I \to \tau\tau) &= \sum_k \sigma(gg\to b\bar{b} h_k) \cdot \mathrm{BR}(h_k\to \tau\tau).\label{Eq:cluster_bbH}
\end{align}
The cluster mass, $m_I$, is determined by a signal strengths weighted mass average
\begin{align}
m_I &= \frac{\sum_k \left[\sigma(gg\to h_k)+\sigma(gg\to b\bar{b}h_k)\right]\cdot \mathrm{BR}(h_k\to\tau\tau) \cdot m_k }{\sum_k \left[\sigma(gg\to h_k)+\sigma(gg\to b\bar{b}h_k)\right] \cdot \mathrm{BR}(h_k\to\tau\tau)}.
\label{Eq:cluster_m}
\end{align}
The sums in Eqs.~\eqref{Eq:cluster_ggH}--\eqref{Eq:cluster_m} run over all Higgs bosons $h_k$ combined in the cluster. In case there is no $h_j$ that fulfills Eq.~\eqref{Eq:combination} for a given $h_i$, the cluster is formed solely by the Higgs boson $h_i$.
It should be noted that taking the incoherent sum of the contributions of the different Higgs bosons involves an approximation. While it is exact in the case of two different $\cp$ eigenstates, e.g.\ $A$ and $H$ in the MSSM, in general interference contributions can be
important~\cite{Fuchs:2014ola,Fuchs:2014zra}. An extension of \HB\ that enables the implementation of interference effects of nearby resonances in a generalized narrow-width approximation is currently under development, see also \citere{ElinaPhD}.

\item In the second step, the expected and observed likelihood values, $q_\text{model}^\text{exp}$ and $q_\text{model}^\text{obs}$, respectively, for each Higgs cluster $h_I$ are evaluated from the experimental likelihood data grid. The likelihood is first evaluated for the rate values $\sigma(gg\to h_I \to \tau\tau)$ and $\sigma(gg\to b\bar{b} h_I \to \tau\tau)$, obtained through Eqs.~\eqref{Eq:cluster_ggH}--\eqref{Eq:cluster_bbH}, in the mass-neighboring data slices, i.e.~at the nearest grid mass values below and above $m_I$, denoted by $m_-$ and $m_+$, respectively. The likelihood value at the predicted cluster mass $m_I$ is then obtained through linear interpolation:
\begin{align}
q(h_I) = \frac{q_-\cdot (m_+ - m_I) + q_+\cdot (m_I - m_-)}{m_+ - m_-}.
\label{Eq:qextrapol}
\end{align}
Here $q_{+/-}$ denote the values of the test statistics obtained at
the neighboring grid above or below the predicted mass $m_I$ (we omitted the subscript `model' for simplicity in Eq.~\eqref{Eq:qextrapol}). These are
obtained, in each case, through bilinear interpolation within the
two-dimensional likelihood planes of the provided CMS data. 

\item The steps 1.~and 2.~are repeated until all $N$ neutral Higgs bosons
have been evaluated as part of at least one Higgs cluster. 
\item Once all likelihoods have been evaluated, the most sensitive analysis application is determined from the resulting \emph{expected} likelihood, $q_\text{model}^{\text{exp}}$, i.e.~the cluster $h_I^\text{max}$ is selected for which $q_\text{model}^\text{exp}(h_I)$ is maximal. The \emph{observed} exclusion likelihood, $q_\text{model}^\text{obs}$, is then used only for this cluster, and provides the final result.
\end{enumerate}
Following this algorithm, the full likelihood from the CMS $\phi \to
\tau\tau$ analysis for both the expected and observed exclusion can be
directly obtained within \HB\ for any tested model. This is carried
out via \texttt{Fortran} subroutines. For a technical documentation see
Appendix~\ref{Sect:Routines}.

The use of this likelihood information is complementary to the other
type of information contained in a full \HB\ application, which
considers exclusion limits from many other Higgs searches from the LEP,
Tevatron and LHC experiments. As an alternative to using the full
likelihood, we therefore also provide the option to reconstruct a limit at
$95\%$ C.L.~and use this in the ``standard'' \HB\ operation. For clarity, we
now repeat some elements of how this works~\cite{Bechtle:2013wla}:
In the statistical procedure, \HB\ first determines the \emph{most sensitive} analysis to the model by picking the analysis application, for which the ratio between the model-predicted signal rate, $S_\text{predicted}$, over the \emph{expected} upper limit on the signal rate, $S^{95\%\text{CL}}_\text{expected}$, 
\begin{align}
r_\text{expected} \equiv \frac{S_\text{predicted}}{S^{95\%\text{CL}}_\text{expected}},
\label{Eq:r_exp}
\end{align}
is maximized. After the most sensitive analysis has been determined, the model prediction is confronted with the \emph{observed} exclusion limit of this particular analysis, $S^{95\%\text{CL}}_\text{observed}$. The model is considered to be excluded at the $95\%\CL$, if
\begin{align}
r_\text{observed} \equiv \frac{S_\text{predicted}}{S^{95\%\text{CL}}_\text{observed}} > 1.
\label{Eq:r_obs}
\end{align}

For the CMS $\phi \to \tau\tau$ analysis described above,
$S^{95\%\text{CL}}_\text{expected}$ and $S^{95\%\text{CL}}_\text{observed}$
are a priori not known and need to be determined from the implemented
likelihood distribution. In a numerical procedure, we therefore scale the
model-predicted $gg\to\phi\to\tau\tau$ and $gg\to b\bar{b}\phi\to\tau\tau$
rates with a \emph{universal} factor $\mu$ until the obtained
expected/observed likelihood $q_\mu^\text{exp/obs}$ values are equal
to $5.99$,
corresponding to the two-dimensional $95\%\CL$
interval.\footnote{Technically, allowing for finite numerical precision we
check for equality within $\lesssim 1\%$.} The so-obtained scale factors,
$\mu^\text{exp/obs}_{95\%\mathrm{CL}}$, are then identified with the
expected/observed $95\%\CL$ upper limits on the signal rate, respectively,
which enter Eqs.~\eqref{Eq:r_exp} and~\eqref{Eq:r_obs}. In this way, the
likelihood-based results from the CMS $\phi \to \tau\tau$ analysis can be
incorporated in the standard \HB\ run. 
%%%%%%%%%%%%%%%%%%%%%%%%%%%%%%%%%%%%%%%%%%%%%%%%%%%%%%%%%%%%%%%%%%%%%%%%%%%%%%%
%%%%%%%%%%%%%%%%%%%%%%%%%%%%%%%%%%%%%%%%%%%%%%%%%%%%%%%%%%%%%%%%%%%%%%%%%%%%%%%

\section{Validation}
\label{sec:validation}

Besides the nearly model-independent limits, CMS has also presented model-specific interpretations of their search results.
This has been done for the MSSM, employing the benchmark
scenarios proposed in Ref.~\cite{Carena:2013qia} (see also
Ref.~\cite{Heinemeyer:2013tqa}). Here, we validate our likelihood
implementation in \HB\ against the CMS results for three of these
scenarios: The $m_h^\text{max}$, the \emph{light stop} and the
\emph{low-$M_H$} scenarios (see Ref.~\cite{Carena:2013qia} for
details). The comparison of the reconstructed $95\%\CL$
exclusion line with the official CMS result provides a non-trivial
test of our implementation: Firstly, it checks whether the
exclusion likelihood agrees over a wide range of different compositions
of the gluon fusion and $b$ quark associated Higgs production rates obtained in the
MSSM parameter space, which are mapped onto the two-dimensional likelihood grids (for
fixed Higgs mass) in our reconstruction. Secondly, it
tests whether our simple criterion of combining signal rates of Higgs
bosons which have similar masses (overlapping within $20\%$) is a
reasonable approximation.  Thirdly, the validation
also tests whether the results obtained from the statistical hypothesis
test of a single narrow resonance model can be mapped reasonably well
onto the full neutral Higgs spectrum of the MSSM (and beyond). 

Some deviations can be
expected at the transition between regimes with different contributing Higgs
combinations. As explained above, the implementation in \HB\ is
based on the CMS results for the single narrow resonance interpretation, and
the contributions of different Higgs bosons of a considered model can only
be combined if their mass differences are such that they would appear as a
single resonance in the CMS search. In contrast, in the dedicated CMS
analyses carried out in specific MSSM benchmark scenarios it was possible to
properly combine the contributions from different Higgs bosons at any given mass constellation
since these have been simulated and tested with their particular masses at every parameter point.
Therefore, the dedicated CMS analysis is expected to have a higher sensitivity than the \HB\ implementation if multiple Higgs bosons with \emph{different} masses each give a non-negligible contribution to the signal yield. Furthermore, due to the simple criterion used in \HB\
for including/excluding the contributions of additional Higgs bosons, the
considered rates in \HB\ may change quite abruptly in a transition region, where the selection of the tested Higgs boson combination changes.
The single resonance approximation is expected to work best when the signal can be described as a single resonance formed by one or several Higgs bosons, while contributions of other Higgs bosons besides those associated with the resonance are negligible.

For predictions in the MSSM benchmark scenarios we employ the $(M_A,\tan\beta)$ grids of Higgs production cross sections and branching fractions for the MSSM benchmark scenarios provided by the LHC Higgs Cross Section Working Group (LHCHXSWG)~\cite{LHCHXSWG_MSSM}.\footnote{The LHCHXSWG cross section and branching fraction grids for the MSSM benchmark scenarios are based on the following set of tools and calculations, that we list here for completeness: \texttt{HIGLU}~\cite{Spira:1995mt}, \texttt{SusHi}~\cite{Harlander:2012pb}, \texttt{FeynHiggs}~\cite{Heinemeyer:1998yj, Heinemeyer:1998np, Degrassi:2002fi, Frank:2006yh,Hahn:2009zz,Hahn:2013ria}, ggH@NNLO~\cite{Harlander:2002wh}, HDECAY~\cite{Djouadi:1997yw,Djouadi:2006bz}, Prophecy4f~\cite{Bredenstein:2006rh,Bredenstein:2006ha}, bbh@NNLO(5FS)~\cite{Harlander:2003ai}, bbh@NLO (4FS)~\cite{Dittmaier:2003ej, Dawson:2003kb}, ggH NLO massive~\cite{Spira:1995rr}, ggH NNLO for scalar Higgs~\cite{Anastasiou:2002yz, Ravindran:2003um}, ggH NNLO for pseudoscalar Higgs~\cite{Harlander:2002vv, Anastasiou:2002wq}, EW corrections from light fermions~\cite{Aglietti:2004nj, Bonciani:2010ms}, (N)NLO (S)QCD corrections for $h/H/A$~\cite{Harlander:2004tp,Harlander:2005rq,Degrassi:2010eu,Degrassi:2011vq,Degrassi:2012vt}.}
For the $gg\to b\bar{b}(h/H/A)$ production process we employ Santander-matching of the $4$- and $5$-flavor scheme (FS) cross sections~\cite{Harlander:2011aa}.

%%%%%%%%%%%%%%%%%% F I G U R E %%%%%%%%%%%%%%%%%%%%%%%%%%%%%%%%%%%%%%%%%%%%%%%%
\begin{figure}
\centering
\subfigure[\label{fig:mhmax_a}~Distribution of the  
observed exclusion likelihood, $q_\text{MSSM}^\text{obs}$, evaluated
with \HB. The contours show the corresponding $95\%\CL$ exclusion limit
(\emph{orange, solid}) and the CMS result obtained from a dedicated
analysis in this scenario~\cite{Khachatryan:2014wca} (\emph{green, dashed}).
]{\includegraphics[width=0.48\textwidth]{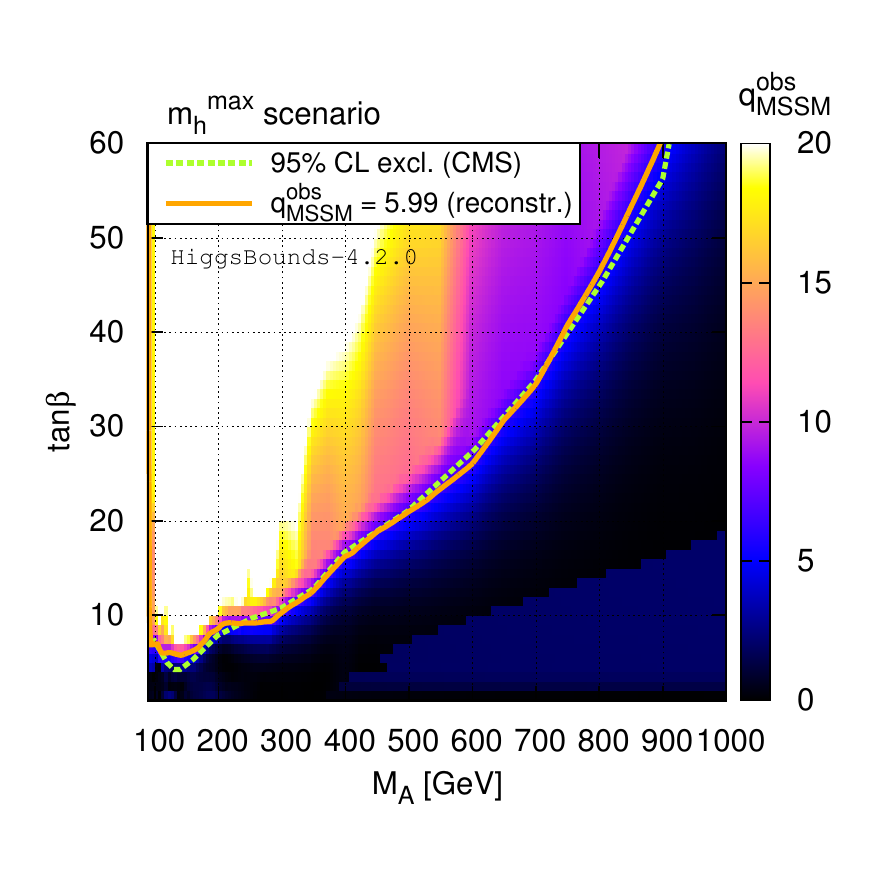}}
\hfill
\subfigure[\label{fig:mhmax_b}~Map indicating the Higgs boson or
cluster of Higgs bosons with the highest sensitivity for a potential exclusion that is used for the likelihood evaluation.]{\includegraphics[width=0.48\textwidth, trim= 0 -0.2cm 0 0]{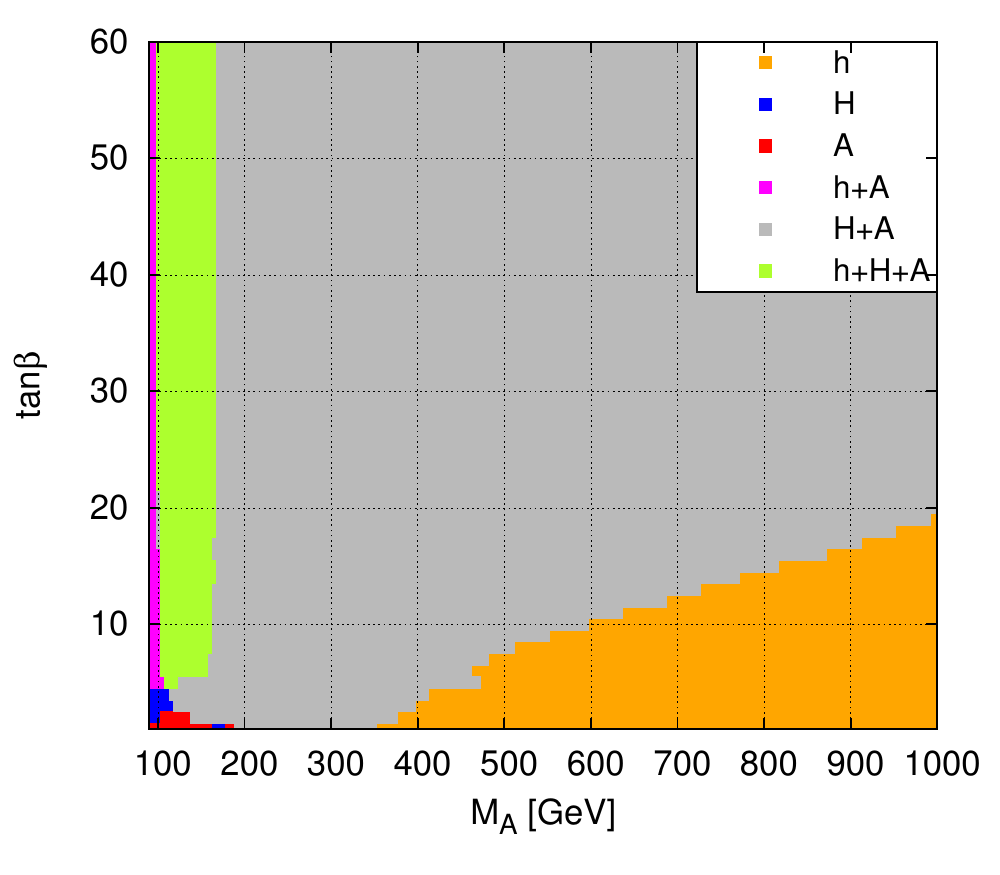}}
\caption{Exclusion likelihood evaluated with \HB\ in the $(M_A, \tan\beta)$ plane of the MSSM $m_h^\text{max}$ scenario.}
\label{fig:mhmax}
\end{figure}
%%%%%%%%%%%%%%%%%% F I G U R E %%%%%%%%%%%%%%%%%%%%%%%%%%%%%%%%%%%%%%%%%%%%%%%%

\newpage
The results for the $m_h^\text{max}$ scenario in the $(M_A, \tan\beta)$
plane are shown in Fig.~\ref{fig:mhmax}.\footnote{Here, and in the following
figures, we show as \HB\ result only the constraints obtained from the CMS
$\phi\to\tau\tau$ analysis and not from the full \HB\ application, where all
currently implemented Higgs searches from LEP, the Tevatron and the LHC
are taken into account.} In
Fig.~\ref{fig:mhmax_a} we show the distribution of the 
observed exclusion likelihood, $q_\text{MSSM}^\text{obs}$ (in color),
as obtained from \HB. 
The corresponding $95\%\CL$ exclusion limit (\emph{orange, solid} contour),
which fulfills $q_\text{MSSM}^\text{obs} = 5.99$, is shown together with the 
CMS result obtained from a dedicated analysis in this benchmark 
scenario~\cite{Khachatryan:2014wca}
(\emph{green, dashed} contour). As mentioned in \refse{sec:expres}, the latter is based on the full combined $7+8\tev$ dataset, whereas the exclusion information implemented in \HB\ is only based on the $8\tev$ dataset. However, this fact is expected to lead to only minor differences in the excluded parameter regions. As can be seen, there is very
good agreement between the exclusion limit reconstructed with \HB\
and the CMS result.
Small deviations can be observed in the low $M_A$ region, $M_A \lesssim
150\gev$, where all three neutral MSSM Higgs bosons contribute substantially
to the signal yield. Here, the result 
reconstructed with \HB\
excludes a slightly
smaller area of parameter space. The \HB\ result can thus be regarded as a
conservative estimate of the actual exclusion limit.

In Fig.~\ref{fig:mhmax_b} we display, for every parameter point in the
$(M_A, \tan\beta)$ plane, the Higgs boson or combination of Higgs
bosons (cluster) that has been selected to obtain the observed exclusion likelihood by the algorithm
described in Section~\ref{Sec:algorithm}. It can be seen that all
three neutral Higgs bosons are combined in most of the parameter region with
$M_A \lesssim 160 - 170\gev$ and $\tan\beta \gtrsim 5$, whereas at larger
$M_A$ values only the two heavier Higgs bosons, $H$ and $A$, which are
nearly mass degenerate, are combined to yield the most sensitive constraint.
At low $\tan\beta$ and large $M_A$ values, however, the combined signal rate
of the heavier Higgs bosons becomes so small that it is instead the light
Higgs boson, with mass around $120-125\gev$, that is selected to give the
observed exclusion likelihood. This is because its \emph{expected} exclusion
likelihood is larger than that obtained for $H/A$. The observed exclusion
likelihood obtained for the light Higgs boson with mass around $125\gev$ is
non-zero because the best-fit point in the two-dimensional cross section
grid at $m_\phi = 125\gev$ is not identical with the SM prediction,
cf.~Fig.~\ref{fig:CMS_BG_m125}. This leads to the small, but non-zero
$q_\text{MSSM}^\text{obs}$ values that are
visible in Fig.~\ref{fig:mhmax_a} at large $M_A$ and small $\tan\beta$.

%%%%%%%%%%%%%%%%%% F I G U R E %%%%%%%%%%%%%%%%%%%%%%%%%%%%%%%%%%%%%%%%%%%%%%%%
\begin{figure}[t!]
\centering
\subfigure[\label{fig:lightstop_a}~Distribution of the observed exclusion likelihood, $q_\text{MSSM}^\text{obs}$, evaluated
with \HB. The contours show the corresponding $95\%\CL$ exclusion limit
(\emph{orange, solid}) and the CMS result obtained from a dedicated
analysis in this scenario~\cite{Khachatryan:2014wca} (\emph{green, dashed}).
]{\includegraphics[width=0.48\textwidth]{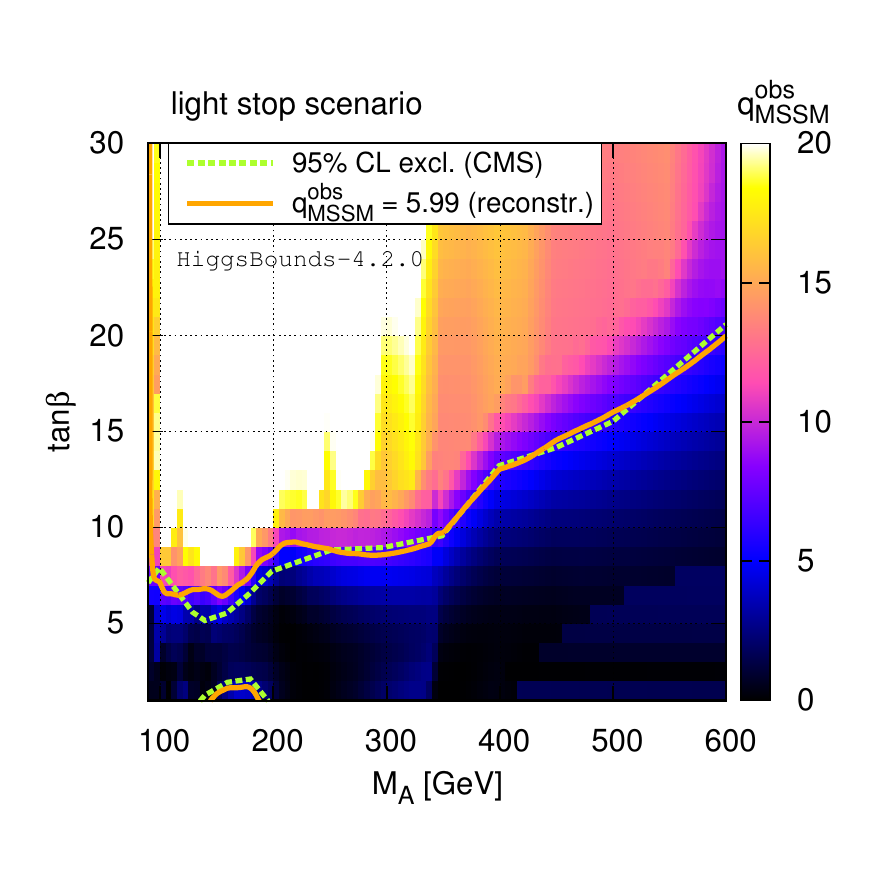}}
\hfill
\subfigure[\label{fig:lightstop_b}~Map indicating the Higgs boson
or cluster of Higgs bosons with the highest sensitivity for a potential 
exclusion that is used for the likelihood evaluation.
]{\includegraphics[width=0.48\textwidth, trim=
0 -0.2cm 0 0]{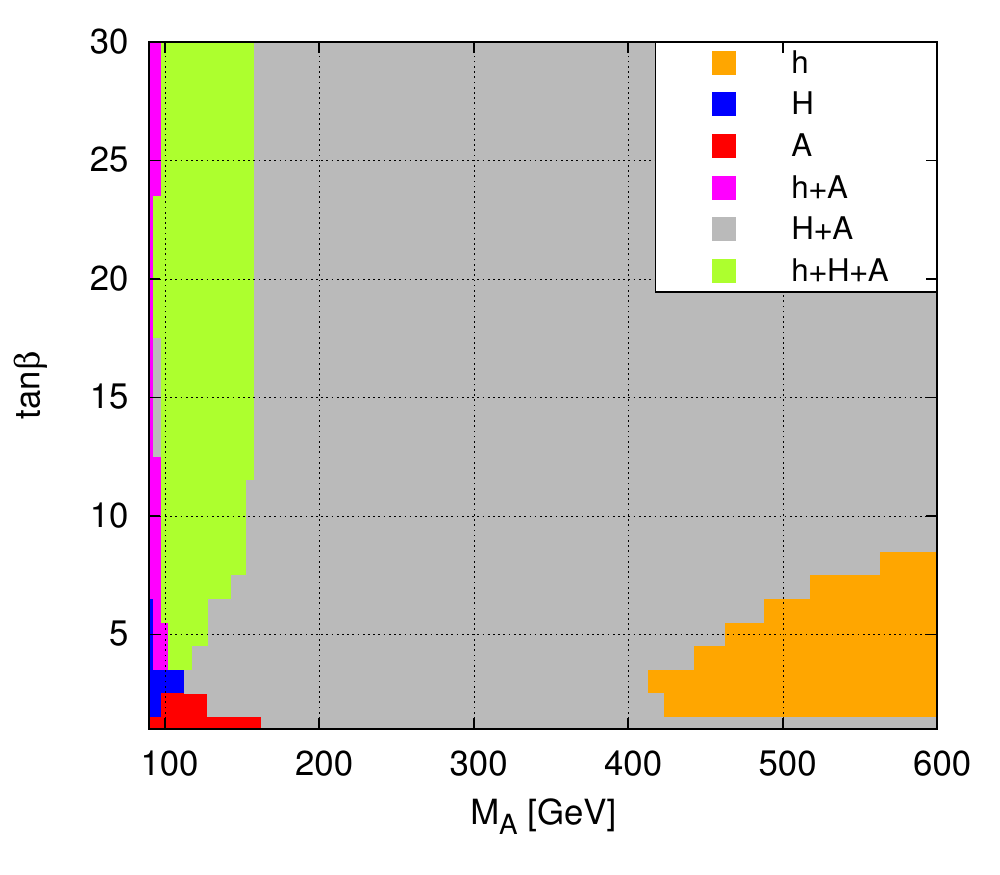}}
\caption{Exclusion likelihood evaluated with \HB\ in the $(M_A, \tan\beta)$ plane of the MSSM \emph{light stop}
scenario.}
\label{fig:lightstop}
\end{figure}
%%%%%%%%%%%%%%%%%% F I G U R E %%%%%%%%%%%%%%%%%%%%%%%%%%%%%%%%%%%%%%%%%%%%%%%%

Next we look at the \emph{light stop} benchmark scenario, for which the cross
section predictions and their associated
theoretical uncertainties have been discussed in detail in
Ref.~\cite{Bagnaschi:2014zla}. This scenario features a
relatively low SUSY particle mass scale, $M_\text{SUSY} = 500\gev$, and
large stop mixing, $X_t=2\, M_\text{SUSY}$, leading to a lightest stop
with a mass of $\sim 325\gev$. This leads to a reduction of
the gluon fusion cross section of the light Higgs by around $10-15\%$
with respect to the SM prediction~\cite{Carena:2013qia}. The 
results of applying our exclusion likelihood implementation in this scenario are shown in
Fig.~\ref{fig:lightstop} (with colors similar to Fig.~\ref{fig:mhmax}). 
The agreement between the 
$95\%\CL$ exclusion contour obtained with \HB\ and the 
CMS result obtained from a dedicated analysis in this benchmark 
scenario~\cite{Khachatryan:2014wca}, as displayed in
Fig.~\ref{fig:lightstop_a}, is very good for pseudoscalar Higgs masses
$M_A \gtrsim 250\gev$. Similarly as in the $m_h^\text{max}$ scenario the
reconstructed exclusion limit obtained from \HB\ 
is slightly weaker for lower $M_A$ values
than the CMS result from the analysis of the benchmark scenarios.
As one can see in Fig.~\ref{fig:lightstop_b}, the reconstructed
likelihood in the low-$\MA$ region obtained from \HB\ is mainly based on a
combination of the $H$ and $A$ signals, while in most part of this parameter
region the light Higgs boson at $125$~GeV is not covered by the $20\%$ mass overlap
criterion used in \HB. In contrast, in the dedicated CMS analysis in
this scenario also the contribution from the light Higgs
boson $h$ is properly combined with the contributions from the other neutral
Higgs bosons. The latter analysis therefore has a slightly higher
sensitivity in this region, which means that the exclusion bound that we
find here is slightly conservative compared to the dedicated CMS result.
In addition to the excluded parameter region at values of
$\tan\beta\gtrsim 5$, the light stop scenario features an additional small
excluded area at lower $\tan\beta$ values, namely $\tan\beta\lesssim 2$,
and $M_A \sim 145 - 190\gev$. The exclusion contour evaluated with \HB\
matches very well with the CMS result also in this region, where gluon 
fusion is the dominant production mode.

Finally, we test our implementation against the results obtained in the
\emph{low}-$M_H$ scenario, where the heavy $\mathcal{CP}$-even Higgs
boson is interpreted as the discovered SM-like Higgs boson at around
$\sim125\gev$ and the light $\cp$-even Higgs is largely decoupled
  from the SM gauge bosons~\cite{Heinemeyer:2011aa}. 
Unlike the other benchmark scenarios, which use $M_A$ as a free parameter, this scenario is defined as a two-dimensional parameter plane in $\tan\beta$ and the Higgsino mixing parameter
$\mu$. The mass of the pseudoscalar Higgs, $M_A$, is fixed to $110\gev$, which leads to a 
lightest Higgs mass that varies mostly between $\sim 80\gev$ and $\sim
105\gev$, but reaching even lower values at very low $\tan\beta$ and very
high $\mu$. 
Since in the MSSM a low value of $M_A$ implies also a light charged
Higgs boson, this scenario served in particular as a benchmark for the
LHC searches for charged Higgs bosons in top quark 
decays. In fact, the parameter space for this scenario in the MSSM is meanwhile essentially
excluded~\cite{TheATLAScollaboration:2013wia,Aad:2014kga,CMS:2014cdp} (see also \citere{Stefaniak:2014roa}). Nevertheless this 
benchmark scenario is still very useful for our validation since all three
neutral Higgs bosons have similar masses and thus contribute non-trivially to the analysis. 

%%%%%%%%%%%%%%%%%% F I G U R E %%%%%%%%%%%%%%%%%%%%%%%%%%%%%%%%%%%%%%%%%%%%%%%%
\begin{figure}[t]
\centering
\subfigure[\label{fig:lowMH_a}~Distribution of the 
observed exclusion likelihood, $q_\text{MSSM}^\text{obs}$, evaluated
with \HB. The contours show the corresponding $95\%\CL$ exclusion limit
(\emph{orange, solid}) and the CMS result obtained from a dedicated
analysis in this scenario~\cite{Khachatryan:2014wca} (\emph{green, dashed}).
]{\includegraphics[width=0.48\textwidth]{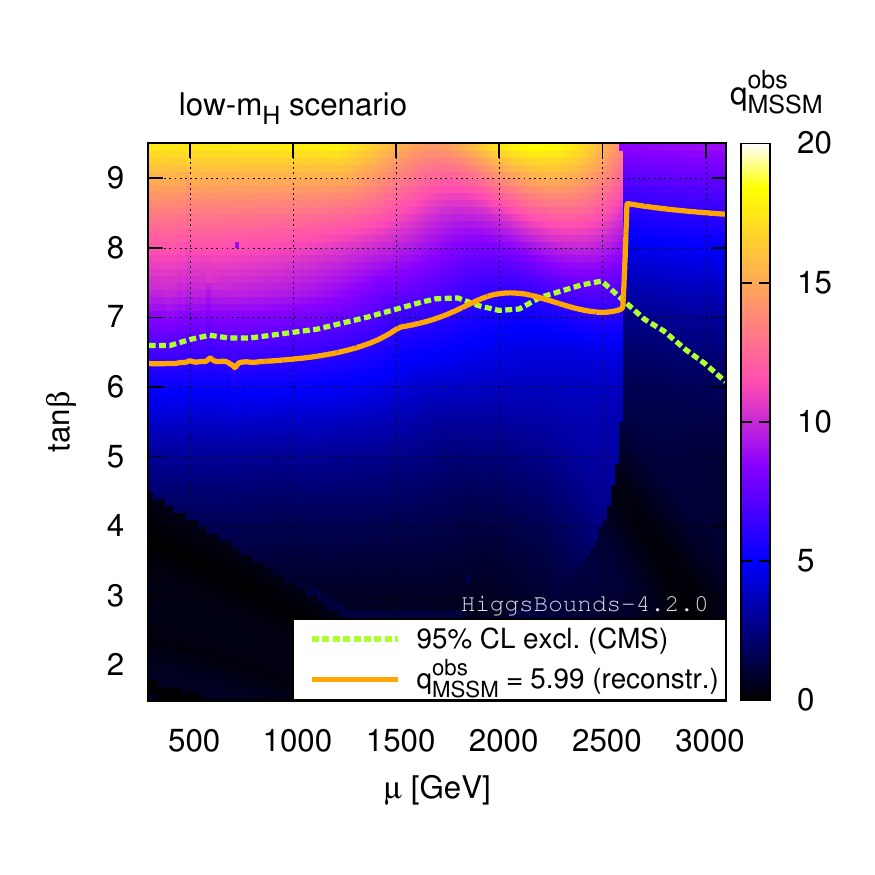}}
\hfill
\subfigure[\label{fig:lowMH_b}~Map indicating the Higgs boson 
or cluster of Higgs bosons with the highest sensitivity for a potential 
exclusion that is used for the likelihood evaluation.
]{\includegraphics[width=0.48\textwidth, trim= 0
-0.2cm 0 0]{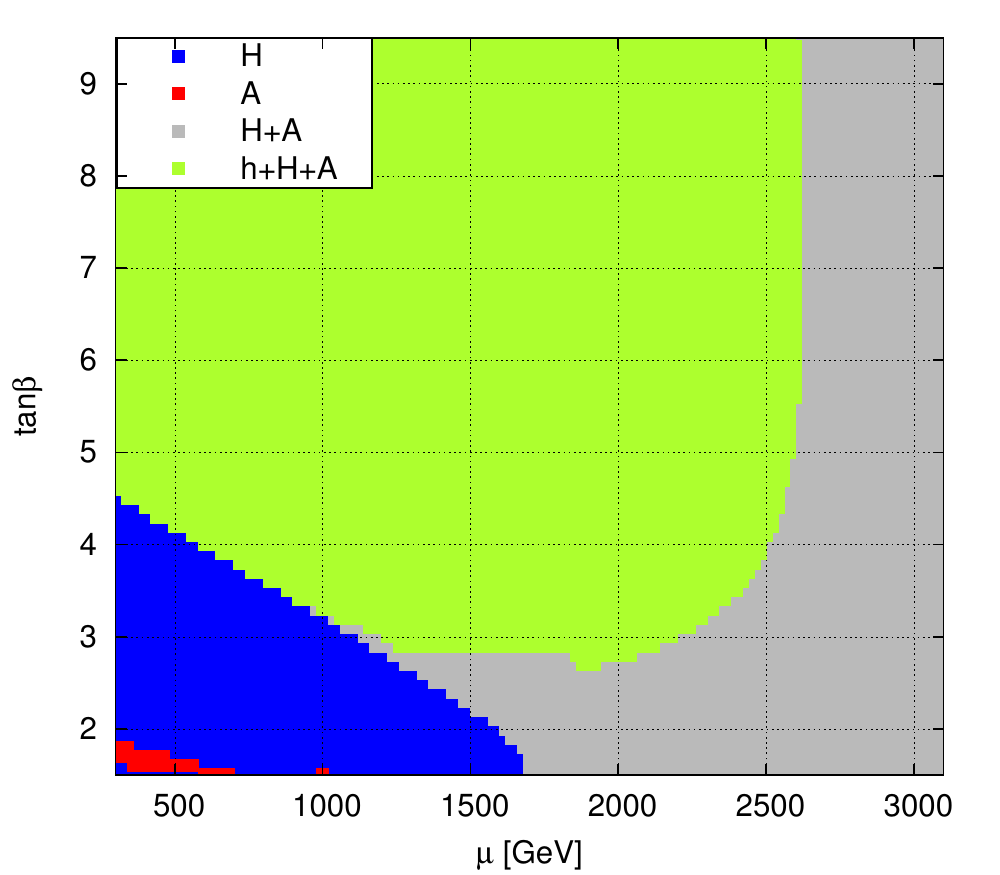}}
\caption{Exclusion likelihood evaluated with \HB\ in the $(M_A, \tan\beta)$ plane of the MSSM \emph{low-$M_H$} scenario.}
\label{fig:lowMH}
\end{figure}
%%%%%%%%%%%%%%%%%% F I G U R E %%%%%%%%%%%%%%%%%%%%%%%%%%%%%%%%%%%%%%%%%%%%%%%%

The comparison of the exclusion limits that have been reconstructed 
with \HB\ with the CMS results is shown in Fig.~\ref{fig:lowMH}. It can be observed in
Fig.~\ref{fig:lowMH_a} that there is rather good agreement between the
exclusion limit obtained with \HB\ and the CMS result for $\mu$ values up to 
$\mu\sim 2600\gev$. At this value of $\mu$ (depending on $\tan\beta$)
the reconstructed exclusion limit develops an ``edge'', and for higher $\mu$
values the reconstructed exclusion limit is significantly weaker than the
one of the CMS result. The reason for this behavior is that at large $\mu$
the light Higgs mass becomes smaller than $88\gev$ and is hence not
combined with the heavier Higgs bosons $A$ and $H$ in \HB. This can be seen in
Fig.~\ref{fig:lowMH_b}. In this parameter region, the tested
signal rate is therefore significantly
smaller than in the case of a full combination of
$h$, $H$ and $A$, and the resulting exclusion limit is accordingly
weaker. In contrast, in the
CMS analysis the signal yield of the light Higgs $h$ has been
properly taken into account even for very low mass values, and
the possibly decreasing signal efficiency is partially compensated by
the increasing production cross section, leading to the significantly
stronger exclusion at high $\mu$ values obtained by CMS.\footnote{A better
  agreement in the large $\mu$ parameter region could be obtained by
  increasing the mass overlap value of $20\%$ in the
  criterion for forming Higgs boson combinations to a sufficiently high value. However, firstly,
  there is no strong physics motivation to choose a value well beyond
  the quoted mass resolution of $\sim20\%$ of the experimental $\tau\tau$ analysis. 
  Secondly, values larger than $20\%$ might lead to too aggressive
exclusions in some scenarios. We therefore stick to the
  value of $20\%$ as the default setting.} 

In almost all the remaining parameter space, all three Higgs bosons
are combined by \HB, as can be seen in Fig.~\ref{fig:lowMH_b},
and the reconstructed exclusion contour resembles 
the CMS result for $\mu$ values below $\mu\sim 2600\gev$. 
The slight deviations observed could result from mass-dependent selection efficiencies for the
$h$ and $H$ signal yields, which cannot be accounted for in the \HB\
implementation since this information is not publicly available. 
Overall, even for this rather extreme scenario in the MSSM we find that
the exclusion likelihood reconstructed with \HB\ approximates the results of
a dedicated analysis reasonably well for large parts of the parameter
space.

%%%%%%%%%%%%%%%%%%%%%%%%%%%%%%%%%%%%%%%%%%%%%%%%%%%%%%%%%%%%%%%%%%%%%%%%%%%%%%%
%%%%%%%%%%%%%%%%%%%%%%%%%%%%%%%%%%%%%%%%%%%%%%%%%%%%%%%%%%%%%%%%%%%%%%%%%%%%%%%

\section{Example application: ``Alignment without decoupling''}
\label{sec:application}

%\begin{comment}

We now go beyond the validation with official CMS results and illustrate the
usefulness of our exclusion likelihood implementation for another MSSM
scenario. We consider here a scenario where the couplings of the light $\CP$-even
Higgs boson become SM-like for a certain range of $\tb$ values, independently of the masses of the remaining Higgs spectrum. The existence of this so-called \emph{alignment limit} was first pointed out in Ref.~\cite{Gunion:2002zf} for the 2HDM.
After the Higgs discovery this possibility has gained attention through a series of
papers~\cite{Craig:2013hca,Asner:2013psa,Carena:2013ooa,Haber:2013mia,Carena:2014nza}, 
see also the ``$\tau$-phobic'' benchmark scenario in \citere{Carena:2013qia}.
In the MSSM the alignment limit can be realized independently of the decoupling of the heavier Higgs states through
a cancellation between tree-level and higher-order contributions in the 
Higgs sector. This cancellation can occur at relatively 
large values of $\tan\beta$ and $\mu \gtrsim M_S$, with $M_S=\sqrt{\mste\mstz}$
being the stop mass scale. In the approximation $\tan\beta \gg 1$, and
taking into account for simplicity only the dominant corrections at one loop, the alignment
condition reads~\cite{Carena:2014nza} 
\begin{align}
\tb &= \frac{\Mh^2 + \MZ^2 +\frac{3 \mt^4 \mu^2}{4\pi^2 v^2  \MS^2}
\left(\frac{\At^2}{2\,\MS^2} -1\right)}
            {\frac{3 m_t^4 \mu \At}{4\pi^2 v^2 M_S^2}
\left(\frac{\At^2}{6\,\MS^2} -1 \right)}.
\label{Eq:alignmentcondition}
\end{align}
Here, $M_Z$ and $m_t$ are the $Z$ boson and top quark mass,
respectively. $M_h$ denotes the light $\cp$-even Higgs boson mass
in the above approximation, and $v\approx 246\gev$. $A_t$ is the trilinear soft-breaking
term in the stop sector.

Solutions of Eq.~\eqref{Eq:alignmentcondition} with $\tb > 0$
exist if $\mu A_t (A_t^2 - 6M_S^2) > 0$. 
Typically, in order to achieve the correct Higgs mass $M_h \sim
125\gev$ for not too large values of the stop masses, the stop mixing is
chosen in the region where the prediction for $M_h$ is maximized,
i.e.~$|X_t| \sim |A_t| \sim \pm\sqrt{6} M_S$ (at
one-loop). Therefore,  for $\mu A_t > 0$ ($ \mu A_t <0$), the alignment
condition has viable solutions for values of $|A_t|$ that are slightly above
(below) the value where the prediction for $M_h$ is maximized.
By increasing $|\mu A_t / M_S^2|$ it is
possible to lower the $\tan\beta$ value at which alignment occurs. 
An MSSM benchmark scenario of this kind for BSM Higgs searches at the 
LHC has recently been proposed in \citere{Carena:2014nza}.

Here, we investigate the benchmark scenario proposed in
Ref.~\cite{Carena:2014nza}, which is essentially a modification of the
$m_h^{\text{mod}+}$ scenario~\cite{Carena:2013qia}
to allow for alignment independent of
decoupling. This so-called $m_h^\text{alt}$ scenario is defined by the
parameter values (in the on-shell scheme) 
\begin{align}
M_1 = 100\gev,\quad M_2 = 200\gev, \quad M_3 = 1500\gev,\nonumber \\
m_{\tilde{\ell}} = m_{\tilde{q}} \equiv M_Q, \quad A_\ell = A_q \equiv A_t,\quad A_t/M_Q = 2.45.
\label{Eq:alignmentbenchmark}
\end{align}

In contrast to the benchmark scenarios of Ref.~\cite{Carena:2013qia},
the parameters $\mu$ and $M_Q$ are adjustable parameters 
in the $m_h^\text{alt}$ scenario.
For convenience, the slepton, sbottom and first and second generation squark soft-breaking
mass parameters are set to $M_Q$, however, these can easily be adjusted to
higher values in order to avoid constraints from SUSY searches at the LHC as
their influence on the Higgs phenomenology is negligible here. We follow the
suggestion
to set $M_Q$ to $1\tev$ per default and, if necessary, increase this
value until a light Higgs mass of $m_h \ge 123\gev$ is obtained. In
practice, this is only relevant at very low values of $\tan\beta$ in the benchmark scenario defined by
Eq.~\eqref{Eq:alignmentbenchmark}. The parameter $\mu$ is then adjusted
according to a chosen ratio $\mu/M_Q$. We focus here on the choice $\mu/M_Q = 3$, implying rather large values of $\mu$, where alignment independent of
decoupling occurs at $\tan\beta\sim 10$.

The MSSM predictions are obtained using the public computer codes
\texttt{FeynHiggs-2.10.2}~\cite{Heinemeyer:1998yj,Heinemeyer:1998np,Degrassi:2002fi,Frank:2006yh,Hahn:2013ria} for the Higgs masses, couplings and
branching fractions, and \texttt{SusHi-1.4.1}~\cite{Harlander:2012pb}
for the gluon fusion and $b$ quark associated production cross sections of the three neutral Higgs bosons. 

%%%%%%%%%%%%%%%%%% F I G U R E %%%%%%%%%%%%%%%%%%%%%%%%%%%%%%%%%%%%%%%%%%%%%%%%

\begin{figure}[t]
\subfigure[\label{Fig:alignmentresults_a}~Likelihood from $h/H/A\to \tau\tau$ exclusion.]{\includegraphics[width=0.49\textwidth]{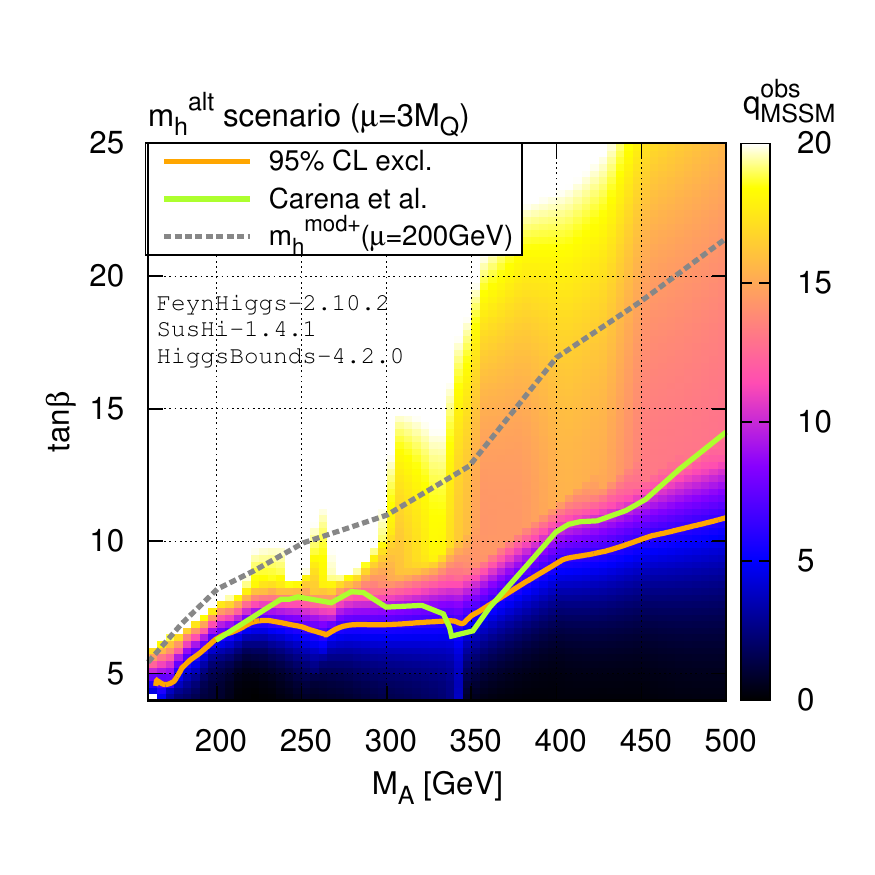}}
\hfill
\subfigure[\label{Fig:alignmentresults_b}~Likelihood from Higgs signal rates.]{\includegraphics[width=0.49\textwidth]{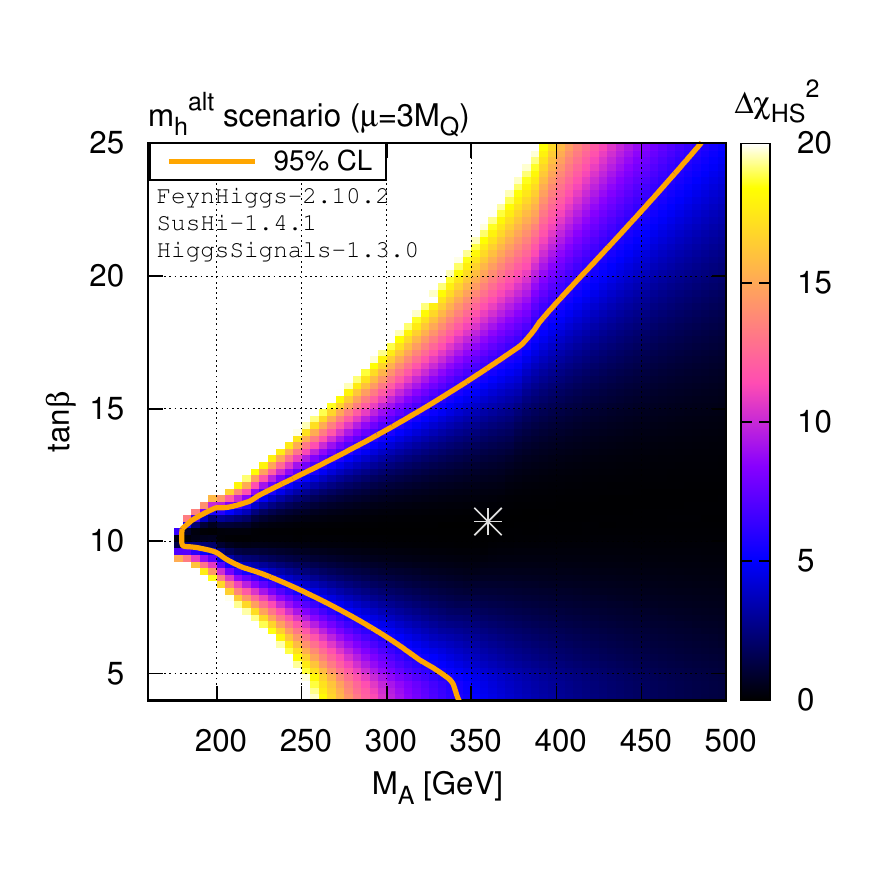}}
\caption{Constraints from LHC Higgs searches in the alignment benchmark
scenario $m_h^\text{alt}$ (with $\mu=3M_Q$): (\emph{a}) Distribution of the
exclusion likelihood from the CMS $\phi\to\tau\tau$ search and 
observed $95\%\CL$ exclusion line as obtained
from \HB. For comparison, also the corresponding $95\%\CL$ exclusion
line given in \citere{Carena:2014nza} (\emph{green, solid}) and the $95\%\CL$ exclusion
line in the $m_h^{\text{mod}+}$ scenario with $\mu = 200\gev$ obtained from
\HB\ (\emph{gray, dashed}) are shown.
(\emph{b}) Likelihood distribution, $\Delta\chi^2_\text{HS}$, obtained
from testing the signal rates of the light Higgs boson $h$ against a
combination of Higgs rate measurements from the Tevatron and LHC
experiments, obtained with \HS. The minimal $\chi^2$ is found at the gray asterisk.}
\label{Fig:alignmentresults}
\end{figure}

%%%%%%%%%%%%%%%%%% F I G U R E %%%%%%%%%%%%%%%%%%%%%%%%%%%%%%%%%%%%%%%%%%%%%%%%

The numerical results for this benchmark scenario are displayed in
Fig.~\ref{Fig:alignmentresults}. The observed exclusion likelihood from
the CMS $\phi\to\tau\tau$ search as obtained from \HB\ is shown in color in
Fig.~\ref{Fig:alignmentresults_a}, and the orange contour indicates the
resulting observed $95\%\CL$ exclusion line. For comparison,
the green contour shows the exclusion line obtained in 
\citere{Carena:2014nza} using results from the same CMS analysis,
however, following a more simplistic approach.\footnote{The limit in
Ref.~\cite{Carena:2014nza} has been
  obtained by ``reverse-engineering'' an inclusive $[\sigma(gg\to\phi) +\sigma(gg\to b\bar{b}\phi)] \times \text{BR}(\phi\to\tau\tau)$ limit from the CMS results for the $m_h^{\text{mod}+}$
  scenario with $\mu=200\gev$~\cite{Carena:2013qia}, and applying this cross section limit
  to the given alignment benchmark scenario. 
In particular, this approximation does not take into account 
the sensitivity of the limit on the individual
  contributions from the gluon fusion and $b$ quark associated Higgs production processes.}
As can be seen from the figure, the more advanced implementation of the observed exclusion likelihood
in \HB\ leads to a somewhat stronger $95\%\CL$ exclusion limit over most of
the parameter space. The relative behavior seems to be different in
the $t\bar{t}$ threshold region, $M_A \approx 2m_t
\approx 345\gev$, where in particular the $gg\to A$ cross section is
enhanced. However, the approximations made in 
Ref.~\cite{Carena:2014nza} appear to be least reliable in this region.
The gray dotted line shows the exclusion limit obtained by
CMS for the $m_h^{\text{mod}+}$ scenario. 
As discussed in \citere{Carena:2013qia}, the excluded regions in the
benchmark scenarios are significantly affected if decay modes of the heavy
Higgs bosons $H$ and $A$ into SUSY particles are kinematically open and
unsuppressed. The presence of such decay modes leads to a sizable reduction of the 
$H/A\to \tau\tau$ branching fractions and therefore to a smaller excluded
region.
In the alignment scenario $\mu$ is very large, leading to a negligible Higgsino component in the
light neutralinos and chargino. The branching fractions for the
Higgs decays to neutralinos and charginos are therefore
essentially absent. In addition, the heavy Higgs decays to gauge bosons,
$H\to W^+W^-$ and $H\to ZZ$, are also suppressed, as the responsible coupling $\propto \cos(\beta-\alpha)$ 
vanishes in the alignment limit. As a result, the $H/A\to\tau\tau$
branching fraction is significantly higher in the alignment scenario 
than in the $m_h^{\text{mod}+}$ scenario, 
which leads to a much larger excluded region in
the alignment scenario, see also the discussion in Ref.~\cite{Carena:2014nza}.

%%%%%%%%%%%%%%%%%% F I G U R E %%%%%%%%%%%%%%%%%%%%%%%%%%%%%%%%%%%%%%%%%%%%%%%%

\begin{figure}[t]
\includegraphics[width=0.49\textwidth]{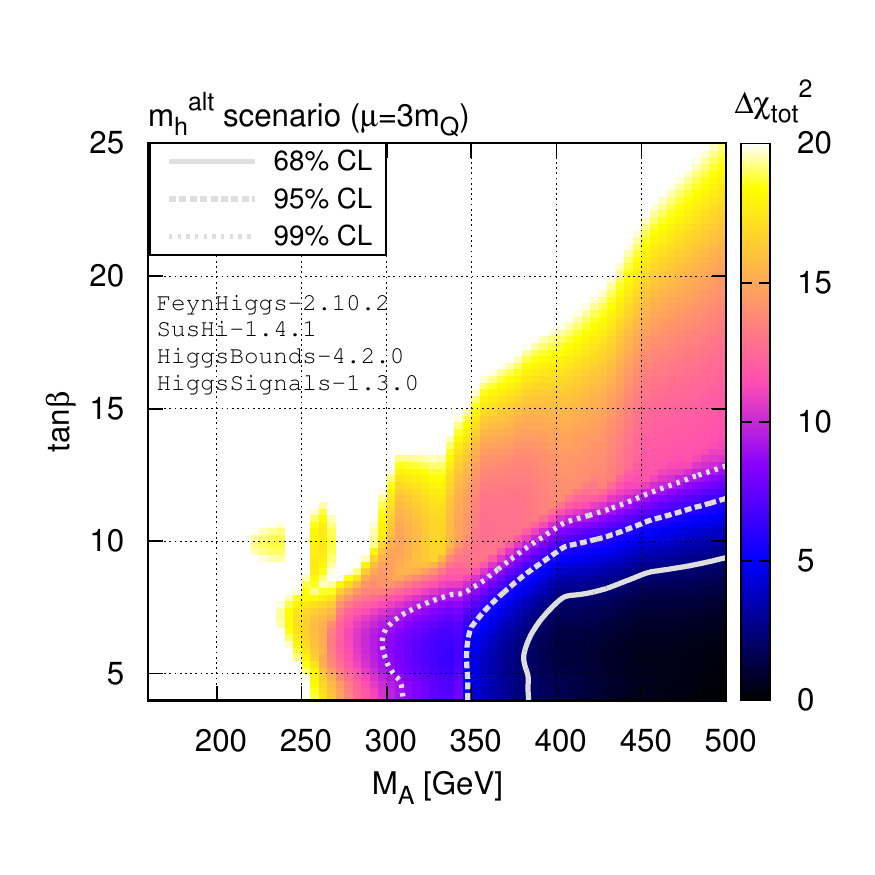}
\caption{Combination of constraints from the CMS $\phi\to\tau\tau$ search and the latest Higgs rate measurements in the MSSM alignment scenario (with $\mu=3M_Q$): The global $\chi^2$ function, $\Delta\chi^2_\text{tot}$, based on the likelihoods provided by \HB\ and \HS, is shown in color; The contours indicate the $1\sigma$, $2\sigma$ and $3\sigma$ allowed regions.}
\label{Fig:alignmentresults_comb}
\end{figure}

%%%%%%%%%%%%%%%%%% F I G U R E %%%%%%%%%%%%%%%%%%%%%%%%%%%%%%%%%%%%%%%%%%%%%%%%

In order to illustrate the complementarity between the constraints from
the CMS $\phi\to\tau\tau$ search and the constraints obtained from the
signal rate measurements of the discovered Higgs boson, we show in
Fig.~\ref{Fig:alignmentresults_b} the likelihood distribution,
$\Delta\chi^2_\text{HS}$, obtained from a $\chi^2$ test of the light
Higgs boson signal rates against a combination of the latest rate
measurements from the
LHC~\cite{Aad:2014eha,Aad:2014eva,ATLAS:2014aga,Aad:2015vsa,Aad:2014xzb,Khachatryan:2014ira,Chatrchyan:2013mxa,Chatrchyan:2013iaa,Chatrchyan:2014vua}
and the Tevatron~\cite{Aaltonen:2013ipa,Abazov:2013gmz}, using the public computer code
\texttt{HiggsSignals-1.3.0}~\cite{Bechtle:2013xfa} (see also Refs.~\cite{Bechtle:2014ewa,Stal:2013hwa}). The
$95\%\CL$ preferred region lies within the orange
contours in Fig.~\ref{Fig:alignmentresults_b}. It is given by the $\chi^2$
difference with respect to the minimal $\chi^2$ value (located in the alignment region and indicated as gray asterisk in Fig.~\ref{Fig:alignmentresults_b}),
$\Delta\chi^2_\text{HS} \equiv \chi^2_\text{HS} - \chi^2_\text{HS,min} \le 5.99$. It can be seen
that  the $\chi^2$ distribution becomes
independent of $M_A$ at around $\tan\beta \approx 10$, indicating that the couplings of the light Higgs
become SM-like independently of the decoupling of the heavier Higgs
states. 

Since we now have
the exclusion likelihood $q_\text{MSSM}^\text{obs}$ from the CMS
$\phi\to\tau\tau$ search available, we can perform a statistical
combination with the constraints from the Higgs rate measurements by
constructing the global $\chi^2$ function $\chi^2_\text{tot} =
q_\text{MSSM}^\text{obs} + \chi^2_\text{HS}$. The resulting
$\Delta\chi^2_\text{tot}$ distribution\footnote{Again,
  $\Delta\chi^2_\text{tot}$ is the $\chi^2$ difference with respect to
  the minimal $\chi^2$ value (obtained at $M_A =500\gev$, $\tan\beta=4$, 
i.e.\ in the lower right corner of Fig.~\ref{Fig:alignmentresults_comb}), now based on the global likelihood
  $\chi^2_\text{tot}$.} is shown in
Fig.~\ref{Fig:alignmentresults_comb}.  The constraints
from the $\phi\to\tau\tau$ searches at the LHC
are highly complementary to the rate measurements,
since they are particularly sensitive at higher values of $\tan\beta$ where the
production process $gg \to b\bar{b} \phi$ is enhanced. In the $m_h^\text{alt}$ scenario with $\mu=3M_Q$,
the combination of both constraints yields a lower limit of $M_A \gtrsim
350\gev$ at the $95\%\CL$ Thus, alignment of the light Higgs
boson occurring without the simultaneous decoupling of the heavier Higgs
states is ruled out for this scenario. 
The \emph{alignment without decoupling} limit can be pushed to lower
values of $\tan\beta$ in this scenario, where the constraints from 
the $\phi\to\tau\tau$ searches are less significant,
only by choosing even more extreme values of $\mu
A_t/M_Q^2$, which potentially leads to problems with vacuum
stability~\cite{Blinov:2013fta}. 

\section{Conclusions}
\label{sec:conclusions}

LHC searches for non-standard Higgs bosons decaying into tau lepton
pairs constitute a sensitive experimental probe for BSM physics.
Recently, the CMS collaboration published the likelihood information
for their Higgs boson searches in the $\tau^+\tau^-$ final
state~\cite{Khachatryan:2014wca}. The likelihood is given as a function
of the two relevant Higgs production channels, gluon fusion and $b$~quark
associated production, for various mass values of the narrow resonance
assumed for the signal model. 
In this paper we have shown how this experimental information can be
utilized to test large classes of theoretical models. In particular, we have developed
a simple algorithm that maps an arbitrary model with multiple neutral Higgs
bosons onto a model with a single narrow resonance, for which the corresponding exclusion likelihood from the CMS search can be determined. We have described the inclusion of this method
into the new version of the publicly available Fortran code \HB\
(version \texttt{4.2.0} and higher).
For nearly any model under consideration, \HB\ provides an evaluation
of the exclusion likelihood for a model parameter point based on the
information from \citere{Khachatryan:2014wca}. Similarly, if requested,
\HB\ can also perform a test of whether or not a given parameter point is
excluded at the $95\%\CL$ based on {\em all} available searches, including the new
$\tau^+\tau^-$ result. The approach to test BSM models with exclusion limits is complementary 
to testing the compatibility of a given model with the observed Higgs signal (and possible future signals 
of additional Higgs bosons). 
The latter kind of information is contained in the
sister program \HS, and both programs can be used together in order to
obtain the combined likelihood information from the search limits and the
Higgs rate measurements. Both codes are available at
\url{http://higgsbounds.hepforge.org}.

We have validated our implementation of the $\tau^+\tau^-$ search
results into \HB\ by comparing the $95\%\CL$ exclusion contours obtained with \HB\ with the 
ones obtained by CMS from dedicated analyses in three Higgs benchmark
scenarios~\cite{Carena:2013qia} in the MSSM. We found very good
agreement in the parameter regions where the sensitivity of the search is
dominated by a single combination of Higgs bosons that can be identified with a single narrow resonance assuming an experimental mass resolution of $20\%$. As expected, the 
largest but still relatively small deviations occur in parameter regions where {\em all} neutral MSSM Higgs
bosons are relatively close in mass and contribute comparably to the signal yield.

As an application, we have discussed the combined constraints
of the $\tau\tau$ search and the rate measurements of the SM-like Higgs
at $125\gev$ in a recently proposed MSSM benchmark scenario, where the
lightest Higgs boson obtains SM-like couplings independently of the
decoupling of the heavier Higgs states.
Here we combined the $\chi^2$ analysis of the rate measurements for the Higgs signal, evaluated with \HS, with the exclusion likelihood from the non-observation in the
$\tau^+\tau^-$ search channel, evaluated with \HB.
We have shown that the combined information yields very significant
constraints on the available parameter space in this scenario and in fact 
disfavors the ``alignment without decoupling region'' in the studied
benchmark model.

We encourage ATLAS and CMS to continue providing their search results 
including the relevant likelihood information. This will greatly facilitate 
the application of the search results for testing BSM models.

%%%%%%%%%%%%%%%%%%%%%%%%%%%%%%%%%%%%%%%%%%%%%%%%%%%%%%%%%%%%%%%%%%%%%%%%%%%%%%
%%%%%%%%%%%%%%%%%%%%%%%%%%%%%%%%%%%%%%%%%%%%%%%%%%%%%%%%%%%%%%%%%%%%%%%%%%%%%%

\section*{Acknowledgments}

We thank Felix Frensch, Andrew Gilbert, Howie Haber, Sasha Nikitenko, 
Alexei Raspereza and Roger Wolf for helpful discussions. In particular, we are
grateful to Felix Frensch, Andrew Gilbert and Roger Wolf for communication
regarding the CMS results of \citere{Khachatryan:2014wca} and for a careful
reading of our manuscript. This work has been supported by the Collaborative Research Center SFB676 of the DFG,
``Particles, Strings and the early Universe", and in part by the European Commission through the ``HiggsTools'' Initial Training Network PITN-GA-2012-316704. 
The work of S.H.\ is supported in part by CICYT (grant FPA 2013-40715-P) 
and by the Spanish MICINN's Consolider-Ingenio 2010 Program under Grant
MultiDark No.\ CSD2009-00064. 
 T.S.\ is supported in part by U.S.~Department of Energy grant number DE-FG02-04ER41286, and in part by a Feodor-Lynen research fellowship sponsored by the Alexander von Humboldt Foundation. 

\appendix

\section*{User guide: How to obtain the exclusion likelihood with \HB}
\label{Sect:Routines}

The exclusion likelihood information for the CMS $\phi\to\tau\tau$
analysis~\cite{Khachatryan:2014wca} is implemented in \HB\ from version 4.2 on. As described in
Section~\ref{Sec:algorithm} this information is used in a standard
\HB\ run to reconstruct the expected and observed $95\%\CL$
exclusion limit, which is then considered alongside all other available
Higgs search limits in the full \HB\ application. This leads to the
global information whether the tested parameter point is allowed or
excluded at the $95\%~\mathrm{C.L}$. Beyond this information the value for the exclusion likelihood for the model parameter point under investigation, $q_\text{model}$, can also be obtained directly via \HB\ Fortran subroutines, enabling the user to incorporate
this information e.g.~in a global parameter fit. In the following we
document the relevant subroutines that make this
information accessible. 

The main routine that runs the algorithm presented in Section~\ref{Sec:algorithm} to obtain the exclusion likelihood is:
\subroutine{HiggsBounds\_get\_likelihood(}{\htdb{\emph{int}~analysisID}, \emph{int}~Hindex, \emph{int}~nc, \emph{int}~cbin, \emph{dble}~M, \emph{dble}~llh, \htdg{\emph{char($*$)}~obspred})}
The (mandatory) input argument\footnote{Here, and in the following, input arguments
  and optional input arguments are highlighted in dark blue and green,
  respectively. The remaining arguments are output values.}
\texttt{analysisID} specifies the analysis for which the likelihood
should be obtained. At the moment, the CMS $\phi\to\tau\tau$
analysis based on the full $8\tev$ dataset (\texttt{analysisID = 3316}) is the only available likelihood, but the framework is easily extendable for future experimental results.
The output values provide information about the selected Higgs boson combination (or \emph{Higgs cluster}):
  \begin{itemize}
  \item[-] \texttt{Hindex} gives the index $i$ of the Higgs boson $h_i$,
which provided the initial seed to form the dominant Higgs cluster
(cf.~Section~\ref{Sec:algorithm}, \emph{item~1}),
  \item[-]  \texttt{nc} gives the
number of Higgs bosons contained in the combination,
  \item[-] \texttt{cbin} is a binary code (bitmask) for the identifiers of the participating Higgs bosons. The binary code is given by summing over $2^{(i-1)}$
for all involved Higgs bosons.\footnote{The indexing of the Higgs bosons is identical to the ordering in which the user chooses to specify the theoretical input for \HB~\cite{Bechtle:2008jh,Bechtle:2011sb,Bechtle:2013wla}.} For example, in the MSSM (with common indexing $h_1 = h$, $h_2 = H$, $h_3 = A$), the
combination $H$+$A$ would give \texttt{cbin = 6}, whereas a 
cluster formed only by the light Higgs $h$ gives  \texttt{cbin
  = 1}.
\end{itemize}
The output value \texttt{M} gives the averaged mass value, calculated according to
Eq.~\eqref{Eq:cluster_m}, at which the likelihood value has been evaluated. The computed value of the likelihood, $q_\text{model}$, is returned as \texttt{llh}.
The final argument is an optional input, \texttt{obspred}, which takes a string value that can be either \texttt{`obs'} or \texttt{`pred'}, specifying whether the
\emph{observed} or \emph{expected} (or \emph{predicted}) exclusion
likelihood should be evaluated, respectively. The default behavior if this argument is not
provided is that the routine returns the observed likelihood, following the algorithm described in Section~\ref{Sec:algorithm}. 

In addition to the main subroutine we provide the following two auxiliary routines, which may be helpful to understand the obtained results.
\subroutine{HiggsBounds\_get\_likelihood\_for\_comb(}{\htdb{\emph{int}~analysisID}, \htdb{\emph{int}~cbin\_in}, \emph{int}~Hindex, \emph{int}~nc, \emph{int}~cbin, \emph{dble}~M, \emph{dble}~llh, \htdb{\emph{char($*$)}~obspred})}
This routine evaluates the likelihood for a specific selection of Higgs bosons that should be considered for the test. Higgs bosons that are \emph{not} available for a possible formation of a Higgs cluster are specified with the input parameter \texttt{cbin\_in}, which is a binary code following the same convention as \texttt{cbin} above. The remaining arguments are the same as above, with the only exception that \texttt{obspred} is now a mandatory input parameter. Among the available Higgs bosons, the routine selects the Higgs combination with the maximal likelihood value and provides the corresponding results. 
\subroutine{HiggsBounds\_get\_likelihood\_for\_Higgs(}{\htdb{\emph{int}~analysisID}, \htdb{\emph{int}~cbin\_in}, \htdb{\emph{int}~Hindex}, \emph{int}~nc, \emph{int}~cbin, \emph{dble}~M, \emph{dble}~llh, \htdb{\emph{char($*$)}~obspred})}
This auxilliary subroutine works in a similar way as above. However, the additional input argument
\texttt{Hindex} forces the routine to consider only Higgs
clusters that contain the specified Higgs boson $h_i$.

In global parameter fits, where both the conventional \HB\ output ($95\%\CL$ exclusion) as well as the likelihood information is used, it is often convenient to deactivate specific analyses during the standard \HB\ run, since these are better described by the likelihood information. In particular, the $95\%\CL$ limits from previous BSM $\phi\to\tau\tau$ searches can be deactivated if instead the CMS $\phi\to\tau\tau$ exclusion likelihood is used. In order to do so, we provide two new subroutines, contained in the \texttt{Fortran} module \texttt{`channels'}.
\subroutine{HiggsBounds\_deactivate\_analyses(}{\htdb{\emph{int(:)}~analysisID\_list})}
This routine should be called \emph{before} the subroutine \texttt{run\_HiggsBounds} in order to deactivate the analyses specified by the \emph{integer array} \texttt{analysisID\_list}. For convenience, the analysis identifiers of the currently implemented LHC $\phi\to\tau\tau$ searches are given in Tab.~\ref{Tab:tautau}. \subroutine{HiggsBounds\_activate\_all\_analyses(}{)}
This subroutine can be used at any time to re-activate all previously deactivated analyses for the succeeding \HB\ run.

\begin{table}
\begin{tabular}{| l | cccc |}
\toprule
\texttt{analysisID} & Experiment & Luminosity and CM-Energy & Additional notes & Reference \\
\hline
3316		& CMS & $19.7~\text{fb}^{-1}$ at $8\tev$ & using $-2\ln{L}$ reconstruction & \cite{Khachatryan:2014wca}\\
2014049	& ATLAS & $19.5-20.3~\text{fb}^{-1}$ at $8\tev$ & profiled limit on $gg\to b\bar{b} \phi$ process & \cite{Aad:2014vgg}\\
20140492	& ATLAS & $19.5-20.3~\text{fb}^{-1}$ at $8\tev$ & profiled limit on $gg\to \phi$ process & \cite{Aad:2014vgg}\\
\botrule
\end{tabular}
\caption{Implemented $95\%\CL$ exclusion limits from LHC searches for BSM Higgs bosons with $\tau\tau$ final states in \HBv{4.2}. The \texttt{analysisID} is used as a unique identifier for an individual analysis and can be used to deactivate/activate them in \HB\ (see text).}
\label{Tab:tautau}
\end{table}

In order to demonstrate the use of these subroutines, we provide the example program \texttt{HBwithLHClikelihood}, included in the \texttt{/example\_programs/} directory of the \HB\ distribution. This program shows how to obtain the observed exclusion likelihood from the CMS $\phi\to\tau\tau$ analysis in the MSSM $m_h^\text{max}$ scenario, such that the user should be able to directly reproduce Fig.~\ref{fig:mhmax}. The example can be compiled by calling \texttt{`make HBwithLHClikelihood'} in the \HB\ main folder, and run from the \texttt{example\_programs} folder by calling \texttt{`./HBwithLHClikelihood'}. Following a successful run, the \texttt{gnuplot} scripts \texttt{`plot\_mhmax\_llh.gnu'} and \texttt{`plot\_mhmax\_llh\_comb.gnu'} in the same folder then reproduce Fig.~\ref{fig:mhmax}.

\bibliography{main}
\bibliographystyle{JHEP}

\end{document}